\documentclass[aps,prd,superscriptaddress,nofootinbib,tighten,twocolumn]{revtex4}
\usepackage{graphicx}
\usepackage{epsfig}  
\usepackage{epsf}    
\usepackage[tbtags]{amsmath}
\usepackage{amsfonts}
\usepackage{filecontents}
\usepackage{float}
\usepackage{subfig}
\usepackage{dcolumn}
\usepackage{bm}
\usepackage{dcolumn}
\usepackage{textcomp}
\usepackage{multirow}
\usepackage{slashed}
\usepackage{float}
\restylefloat{figure}
\usepackage[]{hyperref}
  \hypersetup{
  bookmarks=true,         
  unicode=true,         
  pdftoolbar=true,     
  pdfmenubar=true,     
  pdffitwindow=true,    
  pdfstartview={FitH},    
  pdfsubject={Sterilenus@DUNE},  
  pdfnewwindow=true,     
  pdfcreator={RevTeX},
  colorlinks=true,     
  linkcolor=red,   
  citecolor=blue,    
  filecolor=black,  
  urlcolor=blue,           
  }
\usepackage{hypcap}

\def\nue{{\nu_e}}
\def\anue{{\bar{\nu}_e}}

\newcommand{\eg}{{\it e.g.}}
\newcommand{\ie}{{\it i.e.}}

\newcommand{\beq}{\begin{equation}}
\newcommand{\eeq}{\end{equation}}
\newcommand{\beqa}{\begin{eqnarray}}
\newcommand{\eeqa}{\end{eqnarray}}

\usepackage{cancel}
\usepackage{soul}

\begin{document}
\title{Two-Higgs doublet solution to  the LSND, MiniBooNE  and muon $g-2$ anomalies}
\author{Waleed Abdallah}
\email[]{awaleed@sci.cu.edu.eg}
\affiliation{Harish-Chandra Research Institute, Chhatnag Road, Jhunsi, Allahabad 211019, India}
\affiliation{Department of Mathematics, Faculty of Science, Cairo University, Giza 12613, Egypt}

\author{Raj Gandhi}
\email[]{raj@hri.res.in}
\affiliation{Harish-Chandra Research Institute, Chhatnag Road, Jhunsi, Allahabad 211019, India}

\author{Samiran Roy}
\email[]{samiran@prl.res.in}
\affiliation{Physical Research Laboratory, Ahmedabad - 380009, Gujarat, India}

\begin{abstract}
We show that one of the simplest extensions of the Standard Model, the addition of a second Higgs doublet, when combined with a dark sector singlet scalar, allows us to:  $i)$ explain the long-standing anomalies in the Liquid Scintillator Neutrino Detector (LSND) and MiniBooNE (MB) while maintaining compatibility with the null result from KARMEN, $ii)$ obtain, in the process, a portal to the dark sector, and $iii)$  comfortably account for the observed value of the muon $g-2$. Three singlet neutrinos allow for an understanding of observed neutrino mass-squared differences via a Type I seesaw, with two of the lighter states participating in the interaction in both LSND and MB. We obtain very good fits to energy and angular distributions  in both experiments. We explain features of the solution presented here and  discuss the constraints that our model must satisfy. We also mention prospects for future tests of its particle content.
\end{abstract}


\maketitle


\section{Introduction}
\label{sec1}
The rock-like stability  of the  Standard Model (SM)~\cite{Tanabashi:2018oca} has provided a powerfully reliable framework for both theoretical and experimental  progress in particle physics over many decades. It cannot, however be denied that over this  period, the consistent  agreement of experimental data (in particular from colliders) with its predictions has also been a source of frustration. This is  especially so since there are undeniably strong qualitative reasons, coupled with physical evidence, to expect that there must be physics beyond the  ambit of the SM. These reasons, and this evidence,  include $a)$ Dark Matter (DM)~\cite{Arun:2017uaw,Kahlhoefer:2017dnp,Gaskins:2016cha,  Bertone:2004pz,Feng:2010gw}, $b)$ the observed matter and anti-matter asymmetry in our Universe~\cite{Pascoli:2020swq, Tanabashi:2018oca, Canetti:2012zc}, $c)$ the existence of small but non-zero neutrino mass differences~\cite{PhysRevLett.87.071301,PhysRevLett.81.1562,PhysRevD.88.032002,PhysRevLett.108.191802},  with masses widely different in magnitude from those  of the charged leptons and quarks, and $d)$  the existence, unsupported by compelling physical reasons, of three  families of quarks and leptons with mixings and  a large mass hierarchy. 

Parallelly, albeit on a relatively smaller scale, extremely important experimental  efforts in non-collider settings have supplemented and buttressed the search for new physics. It has gradually become evident that the landscape here is less bleak, and at present one can point to several experiments which report statistically  significant discrepancies with respect to the predictions of the SM. Some anomalous results which have garnered attention and spurred significant activity in an effort to understand their origin are: $a)$ excesses in electron events in short baseline neutrino experiments, which are now in tension with muon neutrino disappearance data~\cite{Maltoni:Talk} if interpreted as oscillation effects involving a sterile neutrino; $b)$  observed discrepancies in the values of the anomalous magnetic moment of the muon~\cite{Miller:2007kk}  and the electron~\cite{Parker:2018vye}; $c)$ a significant excess in the signal versus background expectation in the KOTO experiment~\cite{Ahn:2018mvc} which searches for the decay of a neutral kaon to a neutral pion and a neutrino pair; $d)$ discrepancies with SM predictions in observables related to $B$-decays~\cite{London:2019nlu}; and finally, $e)$ anomalies in the decay of excited states of Beryllium~\cite{DelleRose:2018pgm}.

Our focus in this work is on a subset of results in category $a)$ above.  Specifically, we address the Liquid Scintillator Neutrino Detector (LSND) excess ($\eg$, Ref.~\cite{Aguilar:2001ty}) and  the MiniBooNE (MB) Low Energy Excess (LEE) ($\eg$, Ref.~\cite{MiniBooNE:2020pnu}). In addition to having appreciable statistical significance, they  have withstood scrutiny  by both theoretical and experimental communities over a period of time. It is thus possible that 
these results in particular indicate genuine pointers to new physics, as opposed to un-understood backgrounds or detector-specific effects\footnote{With regard to LSND and MB, which share many similarities in their overall physics goals and parameter  reach ($\eg$, the ratio of oscillation length versus energy of the neutrino beam), we note that such attribution to their results requires two distinct ``mundane" explanations, given that they differ very significantly in backgrounds and systematic errors.}.  The solution proposed here also helps resolve the discrepancy between the measured (see, $\eg$, Ref.~\cite{Bennett:2006fi}) and theoretically predicted ($\eg$, Ref.~\cite{Lindner:2016bgg}) values of the anomalous magnetic moment of the muon.

We show that one of the simplest possible extensions of the SM, the addition of a second Higgs doublet, when acting as a portal to the dark sector, connects and provides an understanding of all three discrepant results mentioned above. Its function as a portal is achieved via its mixing with a dark ($\ie$, SM singlet) scalar. This mixing in the scalar sector allows  heavier dark neutrinos coupled to the singlet scalar become part of the  link between the SM and the dark sector. The dark neutrinos play two additional roles: $a)$ they participate in the interaction that we use to explain the excess events in  LSND and MB;  $b)$ they help generate neutrino masses via a seesaw mechanism. This lends synergy and economy to the model, the specifics of which we give below. It provides excellent fits to both energy and angular event distributions at LSND and MB.

Our paper is organized as follows: Section~\ref{sec2} briefly gives the specifics of  the MB and LSND anomalies  and has a brief discussion of the observed discrepancy in the value of the muon $g-2$. Section~\ref{sec3} describes $i)$ the Lagrangian of our model and its particle content, $ii)$ how the couplings of the additional scalars to fermions arise, and $iii)$  the generation of neutrino masses. Section~\ref{sec4} focusses on the interaction we use to explain the MB and LSND excesses. Section~\ref{sec5} gives our results and provides an accompanying discussion of their important features. Section~\ref{sec6} discusses the constraints on our model. Section~\ref{sec7}  provides a concluding discussion, and indicates possible future tests of the model.
\section{The MB, LSND and muon $g-2$ anomalies}
\label{sec2}
\subsection{Event excesses in MB and LSND}
\label{sec2A}
Two low energy  neutrino experiments, MB~(see \cite{MiniBooNE:2020pnu} and references therein) and LSND~(see~\cite{Aguilar:2001ty}, and references therein), have observed electron-like event excesses. Over time, it has become evident that the results of both  cannot easily be explained within the ambit of the SM. 
 
MB, based at Fermilab,  uses  muon neutrino and anti-neutrino beams produced by 8~GeV protons impinging upon a beryllium target. The  neutrino fluxes peak at around 600~MeV ($\nu_{\mu}$) and around 400~MeV ($\bar{\nu}_{\mu}$). The detector consists of a 40-foot diameter sphere containing  818 tons of pure mineral oil (CH$_2$) and is located 541~m from the target. Starting in  2002, the  MB experiment has up to 2019 collected a total of $11.27 \times 10^{20}$ Protons on Target (POT) in anti-neutrino mode and  $18.75 \times 10^{20}$ POT in neutrino mode.  Electron-like event  excesses of $560.6 \pm 119.6$ in the neutrino mode, and $79.3 \pm 28.6$ in the anti-neutrino mode, with an overall significance of $4.8\sigma$ have been established in the neutrino energy range 200~MeV$< E^{QE}_{\nu} <$ 1250~MeV. Most of the excess is confined to the range $100$~MeV $< {\rm E}_{\rm vis}< 700$~MeV in visible energy, with a somewhat forward angular distribution, and is referred to as the MB LEE. We note $a)$ that all major backgrounds are constrained by \textit{in-situ} measurements, and $b)$ that  MB, due to being a mineral oil Cerenkov light detector, cannot distinguish photons from electrons in the final state.  Additionally,  under certain conditions, MB  could also mis-identify an $e^+e^-$ pair as a single electron or positron.

LSND was  a detector  with 167 tons of mineral oil, doped with scintillator. It employed neutrino and anti-neutrino beams originating from $\pi^-$ DIF as well as $\mu$ decay-at-rest  (DAR). The principal detection interaction was the inverse beta decay process, $\anue + p \rightarrow e^+ + n$.  The detector looked for   Cherenkov and scintillation light associated with the $e^+$ and the correlated and delayed scintillation light from the neutron capture on hydrogen, producing a 2.2~MeV $\gamma$. The experiment observed $87.9 \pm 22.4 \pm 6.0$ such events above expectations at a significance of $3.8\sigma$, over its run span from 1993 to 1998 at the Los Alamos Accelerator National Laboratory. For  reasons similar to those at MB,  LSND lacked the capability to  discriminate a photon signal from those of an $e^+$, $e^-$ or an $e^+e^-$ pair.

 In addition, we mention the KARMEN experiment~\cite{EITEL200289},  which, like LSND and MB,  employed a mineral oil detection medium,  but was less than a third of the size of  LSND. It  had similar incoming proton energy and efficiencies. Unlike LSND, it saw no evidence of an excess.

There have been numerous attempts to understand both of these excesses. A widely discussed resolution involves the presence of sterile neutrinos with mass-squared values of $\sim 1- 10 $~eV$^2$, oscillating to SM neutrinos, leading to  $\anue$ and $\nue$ appearance~\cite{1805.12028}. It is partially  supported  by deficits in $\nu_e$ events in radioactive source experiments and in $\bar{\nu}_e$  reactor flux measurements as well as results from the reactor experiments. However, this explanation for  LSND and MB excesses has had to contend with gradually  increasing tension with  disappearance experiments and is also disfavoured by cosmological data. For recent global analyses, a full set of references and more detailed discussions of these issues, the reader is referred to~Refs.~\cite{1204.5379,1602.00671,1607.00011, 1609.07803, 1703.00860, Dentler:2018sju,Diaz:2019fwt}. 

The tightening of constraints and parameter space for the sterile-active hypothesis has, in turn, led to a large number 
of proposals to explain one or both of the LSND and MB excesses via new physics \cite{hep-ph/0505216, 0902.3802,1009.5536,1210.1519,1512.05357,1602.08766,1708.09548,1712.08019, 1807.09877,1808.02915,1909.08571,Fischer:2019fbw,Dentler:2019dhz, deGouvea:2019qre,Datta:2020auq,Dutta:2020scq,Abdallah:2020biq,Abdullahi:2020nyr}. Many of these scenarios also face a significant number of constraints. For a discussion of these and for related references, we refer the reader to Refs.~\cite{1807.06137,Arguelles:2018mtc, 1810.07185, Brdar:2020tle}. It is, however, fair to say that at the present time, the search for a  compelling and simultaneous explanation of both the LSND and MB anomalies remains a challenge~\cite{Machado:Talk-Neutrino-2020}.
\subsection{The muon $g-2$ anomaly}
\label{sec2B}
The Lande $g$ factor, and its deviation from the tree level value of $2$, represents one of the most precisely measured quantities in the SM. It thus is also an  excellent probe for new physics. Currently there exists a long-standing and statistically significant discrepancy between its measurement \cite{Bennett:2006fi,Brown:2001mga} and the theoretically predicted value, which involves contributions from quantum electrodynamics, quantum chromodynamics and electroweak theory \cite{Miller:2007kk,  Jegerlehner:2009ry,Lindner:2016bgg, Holzbauer:2016cnd}. Specifically,
\begin{equation}
\Delta a_\mu = a_\mu^{\rm meas}-a_\mu^{\rm theory}=(2.74\pm 0.73)\times 10^{-9}.
\end{equation}
There have been many proposals for new physics which provide possible explanations for this discrepancy (for reviews and a full list of  references, see~\cite{Miller:2007kk,Jegerlehner:2009ry,Lindner:2016bgg, Holzbauer:2016cnd}.). Our attempt in this work, details of which are provided in the sections to follow, is related to a class of possible solutions suggested by several authors \cite{Kinoshita:1990aj,Zhou:2001ew, Barger:2010aj, TuckerSmith:2010ra, Chen:2015vqy, Liu:2016qwd, Batell:2016ove, Marciano:2016yhf, Wang:2016ggf, Liu:2018xkx, Liu:2020qgx, Jana:2020pxx} involving a light scalar with a mass in the sub-GeV range and a relatively weak coupling to muons.
\section{The Model}
\label{sec3}
We extend the scalar sector of the SM by incorporating a second Higgs doublet, $\ie$, the widely studied two Higgs doublet model (2HDM)~\cite{Lee:1973iz, Branco:2011iw} in addition to  a dark singlet real scalar\footnote{The introduction of scalars in order to explain one or more of the anomalies at non-collider experiments mentioned in the Introduction is a feature of many recent papers, $\eg$, Refs.~\cite{Kinoshita:1990aj,Zhou:2001ew, Barger:2010aj, TuckerSmith:2010ra, Chen:2015vqy, Liu:2016qwd, Batell:2016ove, Marciano:2016yhf, Wang:2016ggf, Liu:2018xkx, Liu:2020qgx,Dentler:2019dhz,                 deGouvea:2019qre,Jana:2020pxx,Datta:2020auq,Dutta:2020scq,Abdallah:2020biq}. Our model resembles the approach taken in Refs.~\cite{Dutta:2020scq,Abdallah:2020biq}. In particular, it is essentially a more economical version of the model in Ref.~\cite{Abdallah:2020biq}, without an additional $U(1)$.} $\phi_{h'}$. In addition, three right-handed neutrinos help generate neutrino masses via the seesaw mechanism and participate in the interaction described in the next section.

 We  write  the scalar potential $V$ in the Higgs basis $(\phi_h,\phi_H,\phi_{h'})$~\cite{Branco:1999fs,Davidson:2005cw}, with $\lambda_i$ denoting the usual set of quartic couplings
\begin{eqnarray}
&V\!\!=&\!\! |\phi_{h}|^2\left(\frac{\lambda_1}{2} |\phi_{h}|^2+\lambda_3 |\phi_{H}|^2+\mu_1\right) \nonumber\\
&\!\!\!+&\!\!\!\!\!\!|\phi_{H}|^2\left(\frac{\lambda_2}{2} |\phi_{H}|^2+\mu_2\right)+\lambda_4 (\phi_{h}^\dagger\phi_{H})(\phi_{H}^\dagger\phi_{h})\nonumber\\
&\!\!\!+&\!\!\!\!\!\!\phi_{h'}^2\left(\lambda'_2 \phi_{h'}^2+\lambda'_3 |\phi_{h}|^2+\lambda'_4|\phi_{H}|^2+m'\phi_{h'}+\mu'\right)\nonumber\\
&\!\!\!+&\!\!\!\!\!\!\!\bigg{[}\phi_{h}^\dagger\phi_{H}\Big{(}\frac{\lambda_5}{2} \phi_{h}^\dagger\phi_{H} +\lambda_6|\phi_{h}|^2+ {\lambda_7  |\phi_{H}|^2+ \lambda'_5 \phi_{h'}^2}\!-\!\mu_{12}\Big{)}\nonumber\\
&\!\!\!+&\!\!\!\!\!\!\phi_{h'}(m_1|\phi_{h}|^2+m_2|\phi_{H}|^2+m_{12}\phi_{h}^\dagger\phi_{H})+h.c.\bigg{]},
\end{eqnarray}
where
\begin{eqnarray}
\phi_{h}&=&\left( \begin{array}{c}
H^+_1 \\
\frac{v+H^0_1+i G^0}{\sqrt{2}} \\
\end{array} \right)\equiv\cos\beta\, \Phi_1+\sin\beta\,\Phi_2
,\\
\phi_{H}&=&\left( \begin{array}{c}
H^+_2 \\
\frac{H^0_2+i A^0}{\sqrt{2}} \\
\end{array} \right)\equiv-\sin\beta\, \Phi_1+\cos\beta\,\Phi_2,\\
\phi_{h'}&=&H^0_3/\sqrt{2},
\end{eqnarray}
so that $v^2 =v^2_1+v^2_2 \simeq  (246~{\rm GeV})^2$ and $\tan\beta=v_2/v_1$, where $\langle \Phi_i \rangle = v_i/\sqrt{2}$,  $\langle\phi_h\rangle=v$ while $\langle \phi_H\rangle\!=\!0\!=\!\langle \phi_{h'}\rangle$ and $v,\,v_i$ denote vacuum expectation values (VEVs). Here, $|\phi|^2\equiv \phi^\dagger \phi$ and $H_1^+,G^0$ are the Goldstone bosons which give the gauge bosons mass after the electroweak symmetry is spontaneously broken.
The mass matrix of the neutral CP-even Higgses in the basis $\left(H_1^0, H_2^0, H_3^0\right)$ is given by
\begin{equation} \label{CP-Even-MM}
m^2_{\cal H} = \left( 
\begin{array}{ccc}
\lambda_1 v^2 &\lambda_6 v^2  &  0\\ 
\lambda_6 v^2 & \mu_H & m_{12} v/\sqrt{2}  \\ 
 0 & m_{12} v/\sqrt{2}  &\mu_{h'}
 \end{array} 
\right), 
 \end{equation} 
where $\mu_H = \mu_2 +(\lambda_3 + \lambda_4 + \lambda_5)v^2/2$ and $\mu_{h'}=\mu' +\lambda'_3 v^2/2$. Here, we have used the following minimization conditions of the scalar potential $V$:
\begin{equation}
\mu_1 = -\frac{1}{2}\lambda_1 v^2,~~\mu_{12} = \frac{1}{2}\lambda_6 v^2,~~m_1=0.
\end{equation} 
The matrix in Eq.~(\ref{CP-Even-MM}),   $m^2_{\cal H}$, is diagonalized by $Z^{\cal H}$ as follows.
\begin{equation} 
Z^{\cal H} m^2_{\cal H} (Z^{\cal H})^T = (m^{2}_{\cal H})^{\rm diag}\,, 
~~{\rm with}~~
H^0_i = \sum_{j}Z_{{j i}}^{\cal H}h_{{j}}\,,
\end{equation} 
\begin{equation}
{Z^{\cal H}}=\left(
\begin{array}{ccc}
 1 & 0 & 0 \\
 0 & c_\delta & -s_\delta \\
 0 & s_\delta & c_\delta
\end{array}
\right),~~{\tan{2 \delta}=\frac{\sqrt{2}\, m_{12} v}{\mu_{h'}-\mu_H},}
\end{equation}
where $i)$ $s_\delta\equiv \sin \delta$, $c_\delta\equiv \cos \delta$, $(h_1,h_2,h_3)=(h,H,h')$ are the mass eigenstates, $ii)$  $H^0_1\approx h$ is the SM-like Higgs in the alignment limit ($\ie$, $\lambda_6\sim0$) assumed here, and $iii)$ $m^2_h\simeq \lambda_1 v^2$. The  masses of the {extra} CP-even physical Higgs states $(H,h')$ are given by
\begin{equation}
m^{2}_{H,h'}\simeq\frac{1}{2}\left[\mu_H+\mu_{h'}\!\pm\! \sqrt{(\mu_H-\mu_{h'})^2 + 2 m_{12}^2 v^2 }\right]\!.
\end{equation}
{Also,  the charged and CP-odd Higgs masses, respectively, are given by }
\begin{eqnarray}
m^2_{H^\pm}&=&\mu_2+\lambda_3 v^2/2,\\
m^2_{A}&=&\mu_2 +(\lambda_3 + \lambda_4 - \lambda_5)v^2/2.
\end{eqnarray}
In the Higgs basis the relevant Lagrangian ${\cal L}$ can be written as follows
\begin{eqnarray}
{\cal L}\!\!\! &=&\!\!\!\sqrt{2}\Big[(X^u_{ij} \tilde\phi_h\! +\!\bar{X}^u_{ij} \tilde\phi_H) \bar{Q}_L^i  u^j_R
\!+\!(X^d_{ij}  \phi_h\!+\!\bar{X}^d_{ij} \phi_H) \bar{Q}_L^i  d^j_R\nonumber\\
&+&(X^e_{ij}  \phi_h+\bar{X}^e_{ij} \phi_H) \bar{L}_L^i  e^j_R+(X^\nu_{ij}  \tilde\phi_h +\bar{X}^\nu_{ij} \tilde\phi_H) \bar{L}_L^i  \nu_{R_j}\nonumber\\
&+&\frac{1}{\sqrt{8}}{m_{ij} \bar{\nu}^c_{R_i}{\nu_{R_j}}}+\lambda^N_{ij}\bar{\nu}^c_{R_i}\phi_{h'}\nu_{R_j}+h.c.\Big],
\end{eqnarray}
where
\begin{equation}
X^k_{ij}= Y^k_{ij}\, c_\beta +\tilde{Y}^k_{ij}\, s_\beta,~~\bar{X}^k_{ij}= - Y^k_{ij}\,s_\beta +\tilde{Y}^k_{ij}\, c_\beta,
\end{equation}
and $Y^k,\tilde{Y}^k$ are the Yukawa couplings in the $(\Phi_1,\Phi_2)$ basis. We note that $X^k_{ij}$ and $\bar{X}^k_{ij}$ are independent Yukawa matrices.  The fermion masses receive contributions only  from $X^k_{ij}$, since  in the Higgs basis only $\phi_h$ acquires a non-zero VEV while $\langle \phi_H\rangle=0=\langle \phi_{h'}\rangle$, leading to $X^k= {\cal M}_k/v$, where ${\cal M}_k$ are the fermion mass matrices. In this basis, $\bar X^k_{ij}$ are free  parameters and non-diagonal matrices. Hereafter, we work in a basis in which the fermion (leptons and quarks) mass matrices are real and diagonal, where $U_k {\cal M}_k V^\dagger_k=m_k^{\rm diag}$ are their bi-unitary transformations. 

After rotation, one finds the following coupling strengths of the scalars $h,\, h'$ and $H$ with fermions (leptons and quarks), respectively: 
\begin{equation}\label{phi-ll-coupling}
y^{h}_{f} 
\!=\!\frac{m_f}{v},~y^{h'}_{f} 
\!=\!y^f Z_{{3 2}}^{\cal H}\!=\!y^f\, s_\delta,~y^{H}_{f} 
\!=\!y^f Z_{{2 2}}^{\cal H}\!=\!y^f\,c_\delta,
\end{equation}
where $m_f$ are the SM fermion masses and $y^f$ are the diagonal elements of the rotated $\bar X^f$ which are independent of the Yukawa couplings ($y^{h}_{f}=m_f/v$) of the SM Higgs-fermions interactions. $\delta$ manifestly becomes the scalar mixing angle between the mass eigenstates $(H,h')$ and the gauge eigenstates ($H^0_2, H^0_3$).
For neutrinos, we define $n_{R_i}=(U_{\nu_R})_{ij} \nu_{R_j}$ and  $n_{L_i}=(U_{\nu_L})_{ij}\nu_{L_j}$  such that  the  matrices ${\cal M}_\nu(=v X^\nu)$ and $m$ can be diagonalized as follows.
\begin{equation}
U_{\nu_L} {\cal M}_\nu U^\dagger_{\nu_R} =m^{\rm diag}_D,~~U_{\nu_R} m\, U^\dagger_{\nu_R} =m^{\rm diag}_{\nu_R}.
\end{equation}
One can then  define the following matrices:
\begin{equation}
\lambda^n=U_{\nu_R} \lambda^N U^\dagger_{\nu_R},~~~~~~y^\nu=U_{\nu_L} \bar X^\nu U^\dagger_{\nu_R}.
\end{equation}
 The part of the  Lagrangian describing neutrino masses and interactions is then given by
\begin{eqnarray}
{\cal L}_\nu &=& m^{\rm diag}_{D_i}\, \bar n_{L_i} n_{R_i}+\frac{1}{2} m^{\rm diag}_{\nu_{R_i}}\, \bar n^c_{R_i} n_{R_i} + y^\phi_{\nu_{ij}} \bar n_{L_i} n_{R_j} \phi\nonumber\\
&+& (\lambda^{h'}_{N_{ij}} h'+\lambda^{H}_{N_{ij}} H) \,\bar n^c_{R_i} n_{R_j}+h.c.,
\end{eqnarray} 
where $\phi=A,h,H,h'$.

Consequently the  neutrino mass Lagrangian becomes
 \begin{equation} \mathcal{L}_{\nu}^m 
= \frac{1}{2}  \left( \bar{n}^c_{L} ~\bar n_{R}  \right) \left( \begin{array}{cc}
 0& m^{\rm diag}_D \\[0.1cm]   m^{\rm diag}_D &m^{\rm diag}_{\nu_{R}}    \end{array} \right) \left( \begin{array}{c} n_{L} \\  n^c_{R} \end{array}  \right) +h.c., 
 \end{equation}
and  the neutrino mass matrix is given by
\begin{equation}
m_\nu=\left( \begin{array}{cc}
 0& m^{\rm diag}_D \\[0.1cm]   m^{\rm diag}_D &m^{\rm diag}_{\nu_{R}}   \end{array} \right)  \rightarrow~~~ m_\nu^{\rm diag} = {\cal N} m_\nu \,{\cal N}^\dagger. 
\end{equation}
Its  eigenvalues ($m_\nu^{\rm diag}$)  are
\begin{equation}
m^{\rm diag}_{\nu}\simeq{\rm diag}\left\{-m^2_{D_i}/m_{\nu_{R_i}},m_{\nu_{R_i}}\right\},~~(i=1,2,3).
\end{equation}
The neutrino mass matrix $m_\nu$ can be diagonalised, up to  ${\cal O}(m_{D_i}/m_{\nu_{R_i}})$, by the neutrino mixing matrix ${\cal N}$ which can be written, up to corrections of ${\cal O}(m^2_{D_i}/m^2_{\nu_{R_i}})$, as
\begin{eqnarray}
{\cal N}&\simeq&  \left( \begin{array}{cc}
 I-\Theta^2/2 & \Theta\\  -\Theta &  I-\Theta^2/2 \end{array} \right) ,
 \end{eqnarray} 
 where $\Theta_i= m_{D_i}/m_{\nu_{R_i}}$.  The neutrino mass eigenstates (physical states) are given by
 \begin{equation}
 \left( \begin{array}{c} \nu \\  N \end{array}  \right)= \left( \begin{array}{cc}
 I-\Theta^2/2 & \Theta\\  -\Theta &  I-\Theta^2/2 \end{array} \right) \left( \begin{array}{c} n_{L} \\  n^c_{R} \end{array}  \right).
 \end{equation}
For the normal order ($m_{\nu_1} < m_{\nu_2} < m_{\nu_3}$), the two mass squared differences of the light neutrinos determined from the oscillation data are  $\Delta m_{21}^2 = (7.05 - 8.14) \times 10^{-5}$~eV$^2$ and $\Delta m_{31}^2 = (2.41 - 2.60) \times 10^{-3}$~eV$^2$~\cite{deSalas:2017kay}. We have chosen a benchmark point, see Table~\ref{tab}, so that it satisfies these values. Finally,  the part of the  Lagrangian specifying  neutrino interactions is given by
 \begin{equation}
{\cal L}_\nu^{\rm int} \simeq y^\phi_{\nu_{ij}} \bar\nu_i N_j \phi+ (\lambda^{h'}_{N_{ij}} h'+\lambda^{H}_{N_{ij}} H) \bar N_i N_j+h.c.,
\end{equation}  
 where the coupling strengths of the scalars $h',H$ for vertices connecting active and sterile neutrinos, respectively, are as follows.
\begin{equation}\label{phi-vN-coupling}
y^{h'}_{\nu_{ij}} 
=y^\nu_{ij} Z_{{3 2}}^{\cal H}=y^\nu_{ij}\,s_\delta,~y^{H}_{\nu_{ij}} 
=y^\nu_{ij} Z_{{2 2}}^{\cal H}=y^\nu_{ij}\,c_\delta.
\end{equation}
Additionally, the coupling strengths of the scalars $h',H$ for vertices connecting  two sterile states, respectively, are
\begin{equation}\label{phi-vv-coupling}
\lambda^{h'}_{N_{ij}} 
=\lambda^n_{ij} Z_{{3 2}}^{\cal H}=\lambda^n_{ij}\,c_\delta,~\lambda^{H}_{N_{ij}} 
=\lambda^n_{ij} Z_{{2 2}}^{\cal H}=-\lambda^n_{ij}\,s_\delta.
\end{equation}
\begin{table}[h!]
\vspace{-0.5cm}
\begin{center}
 \begin{tabular}{|@{\hspace{0.03cm}}c@{\hspace{0.03cm}}|@{\hspace{0.03cm}}c@{\hspace{0.03cm}}|@{\hspace{0.03cm}}c@{\hspace{0.03cm}}|@{\hspace{0.03cm}}c@{\hspace{0.03cm}}|@{\hspace{0.03cm}}c@{\hspace{0.03cm}}|@{\hspace{0.03cm}}c@{\hspace{0.03cm}}|}
  \hline
  $m_{N_1}$& $m_{N_2}$ & $m_{N_3}$&$y_{u}^{h'(H)}\!\!\times\!\! 10^{6}$ &$y_{e(\mu)}^{h'}\!\!\times\!\! 10^{4}$&$y_{e(\mu)}^{H}\!\!\times\!\! 10^{4}$ \\
  \hline
   85\,MeV  & $130$\,MeV & $10$\,GeV&$0.8(8)$ &$0.23(1.6)$&$2.29(15.9)$ \\
  \hline
    \hline
  $m_{h'}$& $m_{H}$ &$\sin\delta$&  $y_{d}^{h'(H)}\!\!\times\!\! 10^{6}$&$y_{\nu_{\!i 2}}^{h'(H)}\!\!\times\!\! 10^{3}$&$\lambda_{N_{\!12}}^{h'(H)}\!\!\times\!\! 10^{3}$ \\[0.05cm]
  \hline
   17\,MeV  & 750\,MeV &$0.1$& $0.8(8)$&$1.25(12.4)$&$74.6(-7.5)$ \\
  \hline
 \end{tabular}
\caption{Benchmark point used for event generation in LSND, MB and for calculating the muon $g-2$.}
\label{tab}
\end{center}
\vspace{-0.6cm}
\end{table}

Finally, we stress that {\it all} the Yukawa couplings of the light scalars ($h',H$)-fermion interactions ($y^{h',H}_f,~f=\ell,q,\nu_\ell,N$) are free and independent of the Yukawa couplings ($y^{h}_{f}=m_f/v$) of the SM Higgs-fermion interactions.

\begin{figure}[t!]
\center
\includegraphics[width=0.6\textwidth]{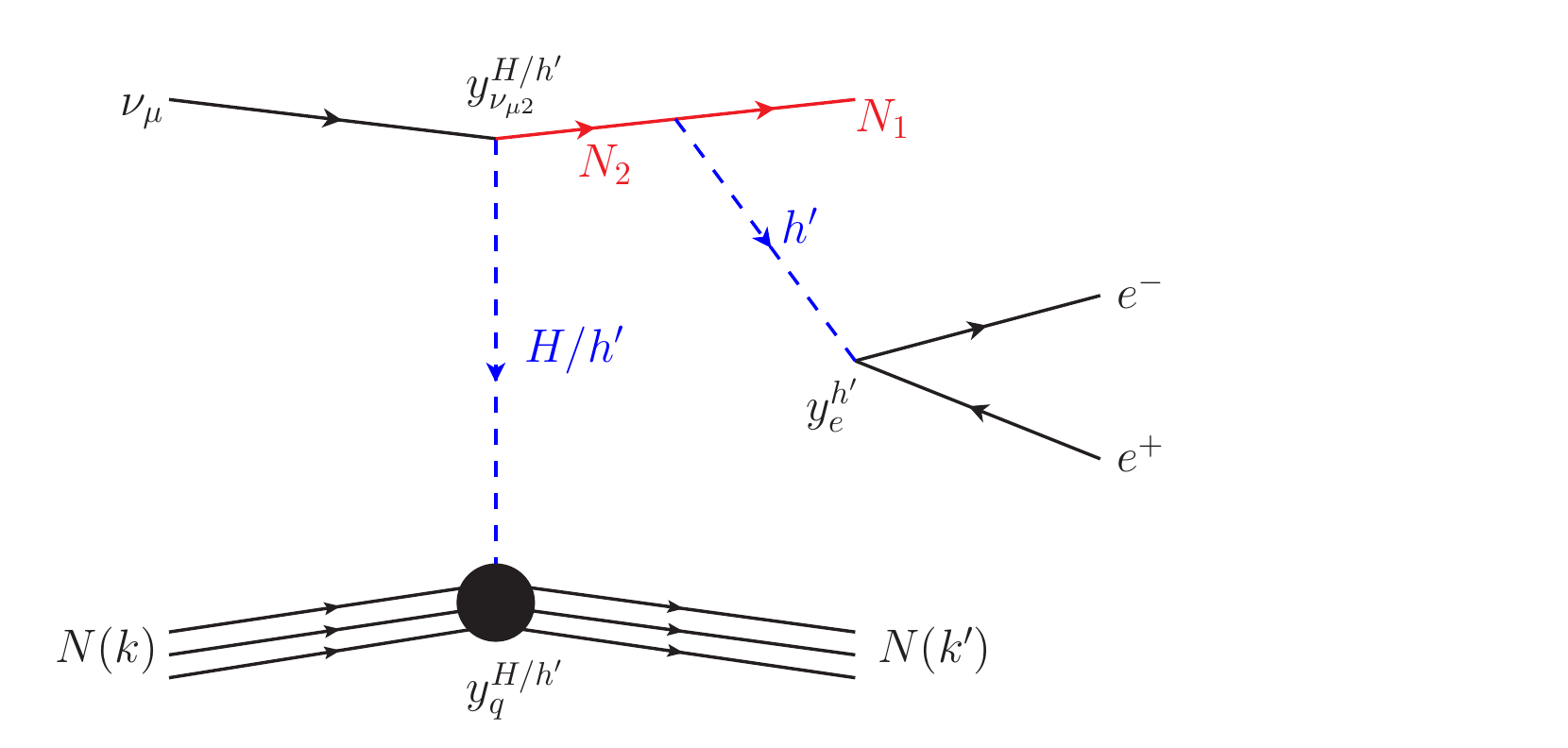}
\caption{Feynman diagram of the  scattering process in our model which leads to the excess in LSND and MB.}
\label{FD-SP-LSND-MB}
\end{figure}

\begin{figure*}[t!]
\includegraphics[width=0.45\textwidth]{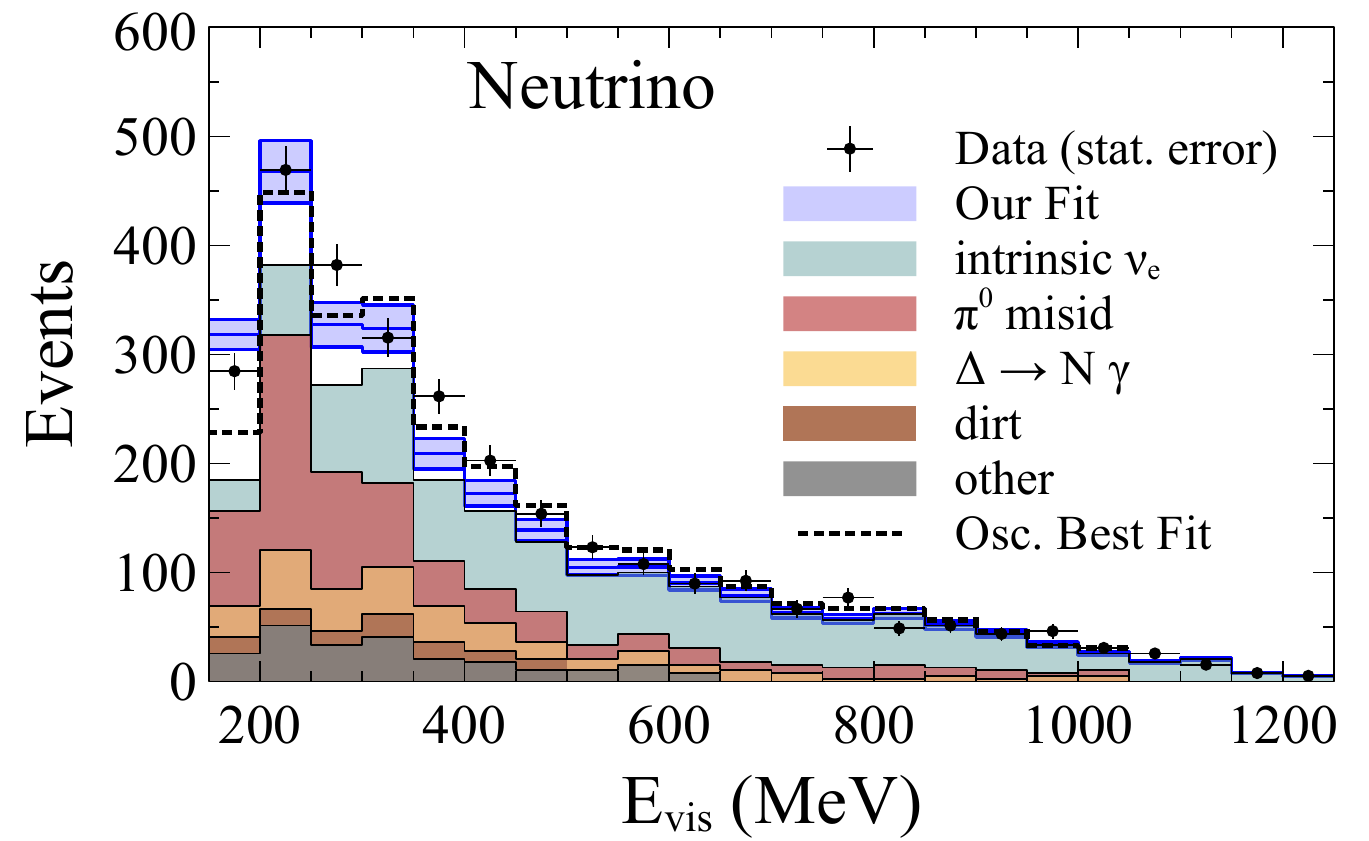}~~
\includegraphics[width=0.45\textwidth]{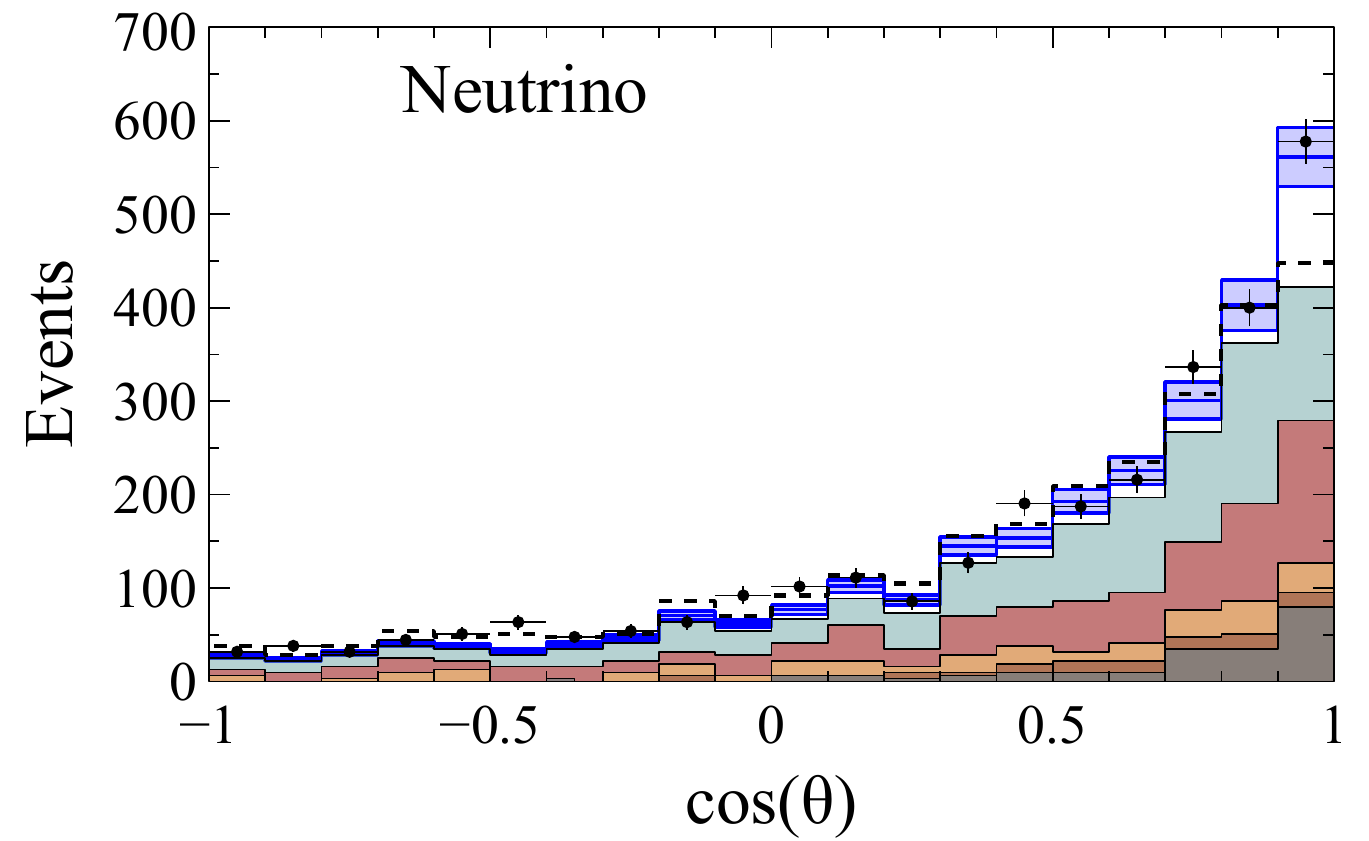}\\[0.2cm]
\includegraphics[width=0.45\textwidth]{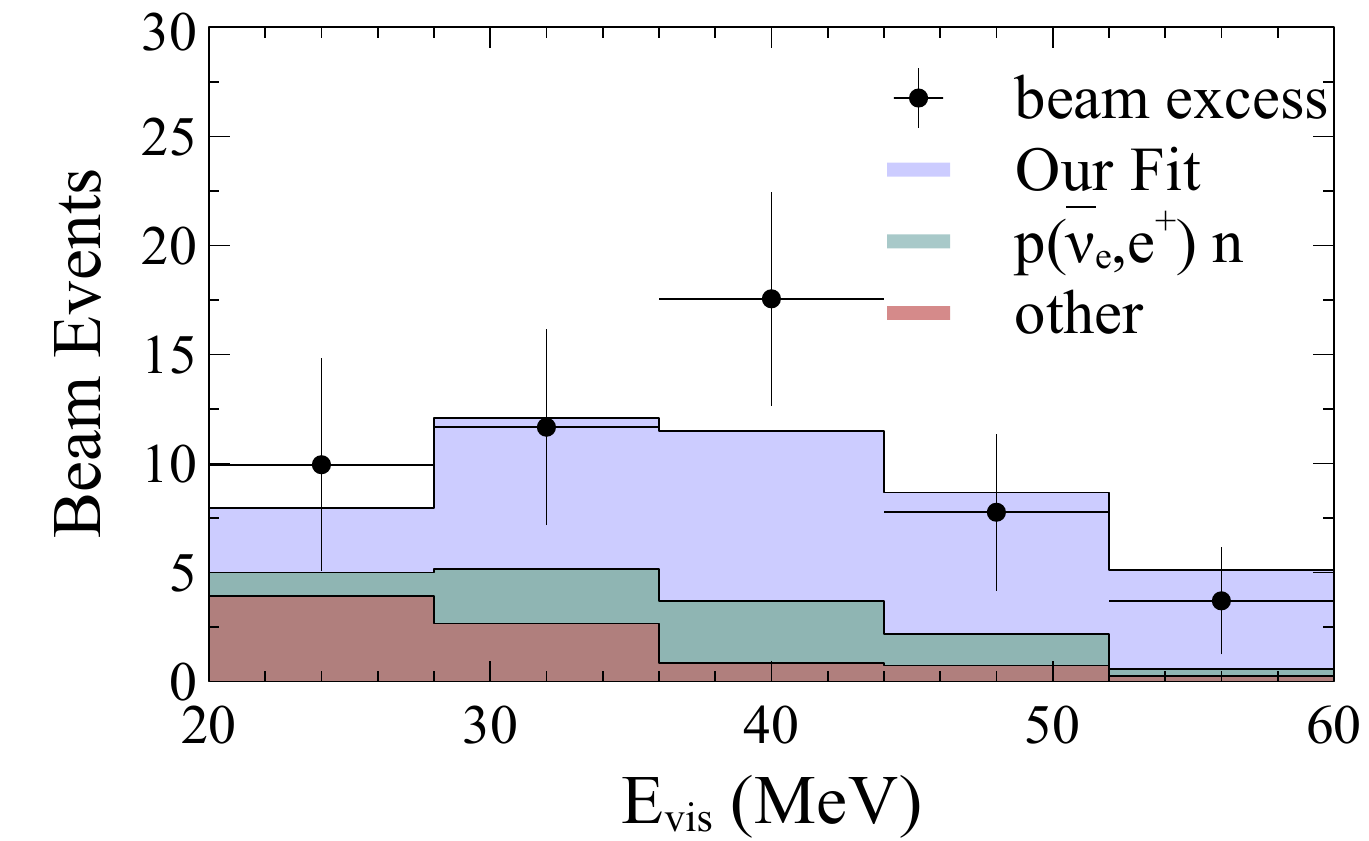}~~
\includegraphics[width=0.45\textwidth]{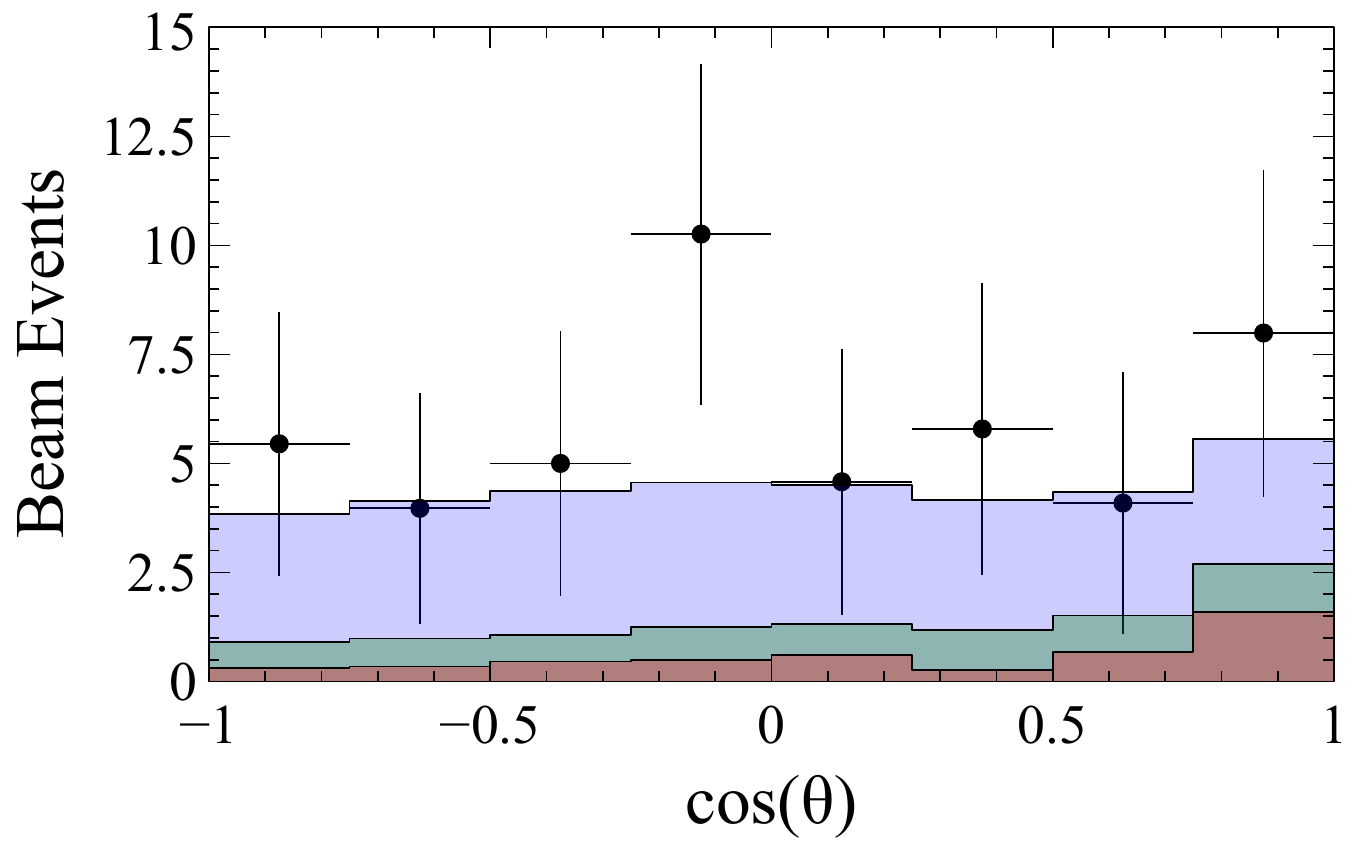}
\caption{Top panels: The MB electron-like events (backgrounds and signal) from~\cite{MiniBooNE:2020pnu}, versus the visible energy  E$_{\rm vis}$ (left panel) and versus the cosine of the emitted angle of the light (right panel), for neutrino runs.  The blue solid line is the prediction of our model. Bottom panels:  The energy distribution (left panel) of the LSND data \cite{Aguilar:2001ty} for $R_\gamma> 10$, and the angular distribution (right panel) of the light due to the electron-like final state, for $R_\gamma > 1$.  The shaded blue region  in both panels is our fit, and other shaded regions are the backgrounds.} 
\label{MB-LSND-events}
\end{figure*}
\section{The interaction in MB and LSND}
\label{sec4}
In the process shown in Fig.~\ref{FD-SP-LSND-MB}, the heavy sterile neutrino $N_2$ is produced via  the upscattering of a muon neutrino ($\nu_\mu =U_{\mu i} \nu_i$) present in the beam,  both for  MB and LSND. Once $N_2$ is produced, it decays promptly to another lighter sterile neutrino $N_1$ and a  light scalar $h'$. In our scenario,  $N_1$ is a long-lived particle that either  escapes the detector or decays to lighter dark particles but $h'$ decays promptly to a collimated $e^+e^-$ pair and produces the visible light that comprises the signal. 

As shown in Fig.~\ref{FD-SP-LSND-MB}, both $H$ and $h'$ act as mediators and  contribute to the total cross section. The contribution of $h'$ is much smaller $(\sim 10\%)$ compared to  that of $H$,  since  $\sin \delta \simeq 0.1$. However,  this small contribution plays an important role in  producing  the correct angular distribution in MB,  as we discuss later. In our model, $H$ and $h'$ predominantly couple to the first generation of quarks ($u$ and $d$) and have negligible or tiny  couplings to other families. The effective coupling $(F_N)$ of either scalar to a nucleon $(N)$ can be written as~\cite{Junnarkar:2013ac,Crivellin:2013ipa,Hoferichter:2015dsa}
\begin{eqnarray}
\dfrac{F_N}{M_N}= \sum_{q=u,d} \, f^{N}_{T_q} \dfrac{f_q}{m_q}.
\end{eqnarray}
Here $M_N$ is the nucleon mass and the values of $(f^{p}_{T_u},f^{p}_{T_d},f^{n}_{T_u},f^n_{T_d}) =( 0.020,0.041, 0.0189,0.0451)$. In our scenario,  $f_{q} = y_{q}^{H,h'}$, $(q=u,d)$.

We include both the incoherent and coherent contribution in the production of $N_2$ in MB.  For LSND, however,  we consider only incoherent scattering from neutrons. The total differential cross section, for the target in MB,  $\ie$, CH$_2$, is given by
{\selectfont\fontsize{9.5}{9.5}{
\begin{equation*}
\Big[\dfrac{d\sigma}{dE_{N_2}}\Big]_{{\rm CH_{2}}}\!=\!\Big[\underbrace{(8F_{p}\!+\!6F_{n})}_{\textrm{\footnotesize{{incoherent}}}}
+\underbrace{ (6F_{p}\!+\!6F_{n})^{2}  e^{-2b|q^{2}|}}_{\textrm{\footnotesize{{coherent}}}}\Big]\dfrac{d\sigma}{dE_{N_2}}.
\label{tot_xsec}
\end{equation*}}}
 The entire  carbon nucleus (C$^{12}$) contributes in coherent scattering, with, however, decreasing contributions as  $|q^2| = |(k'-k)^2| $ increases.  To implement this, we employ a  form factor,  $exp(-2b|q^{2}|)$~\cite{Hill:2009ek}. Here  $b$ is a numerical parameter, which in the case of  C$^{12}$ takes the value  $25$~GeV$^{-2}$~\cite{PhysRevD.9.1389, Hill:2009ek}. The number of events is given by
\begin{equation}
{\rm N}_{\rm{events}} \! =\! \eta\!\! \int\!\! dE_{\nu} dE_{N_2}\dfrac{d\Phi^{\rm{\nu}}}{dE_{\nu}} \!\,  \dfrac{d\sigma} {dE_{N_2}}\! \times\! {\rm BR}(N_2 \!\rightarrow\! N_1 h'),
\end{equation}
with $E_{h'}\in [E_{h'},E_{h'}+\Delta E_{h'}]$ and $\Phi^{\rm{\nu}}$ is the incoming muon neutrino flux.  $\eta$ contains all detector related  information like efficiencies, POT etc. All calculations for LSND, MB and the value of  the muon $g-2$ are carried out using the benchmark values in Table~\ref{tab}.
Finally, for these  values, the calculated  lifetimes of $N_2$ and $h'$ in the rest frame are $10^{-17}$~s and $1.8\times 10^{-12}$~s, respectively.

 Our results are presented in the next section.
\vspace{-0.4cm}
\section{Results and Discussion}
\label{sec5}
\vspace{-0.4cm}
In this section we present the results of our numerical calculations,  using the   cross section for the process and the model described in Section~\ref{sec3}.
\vspace{-0.5cm}
\subsection{Results and discussion for  MB and LSND}
\label{sec5A}
\vspace{-0.4cm}
Fig.~\ref{MB-LSND-events} (top panels) shows the MB data points, SM backgrounds and  the prediction of our model (blue solid line) in each bin. Also shown (black dashed line) is the oscillation best fit. The latest data set  for the neutrino mode, corresponding to $18.75 \times 10^{20}$ POT, as detailed in \cite{MiniBooNE:2020pnu} has been used in our fit. The left  panel shows the distribution of the measured visible energy, ${\rm E}_{\rm vis}$, plotted against the events for neutrinos. In our model, ${\rm E}_{\rm vis}$ is the same as  $E_{h'}$. The  angular distributions for the emitted light are shown in the right panel. The fit corresponds to benchmark parameter values  shown in Table~\ref{tab}. We have used  fluxes, efficiencies, POT exposures, and other relevant information from \cite{Aguilar-Arevalo:2018gpe} and references therein to prepare these plots. We see that very good fits to the data are obtained  for both the  energy and the angular distributions. The data points show only statistical uncertainties. We have assumed a $15\%$ systematic uncertainty for our calculations. These errors are represented by the blue bands in the figures.

As mentioned earlier, the LSND observations measure the visible energy from the Cerenkov and scintillation  light of  an assumed electron-like event,  as well as the $2.2$ MeV photon resulting from coincident neutron capture on hydrogen. In our model, this corresponds to the  scattering diagrams in Fig.~\ref{FD-SP-LSND-MB} where the target is a neutron in the  Carbon nucleus. Unlike the case of MB above, where both coherent and incoherent processes contribute to the total cross section, the LSND cross section we have used  includes only an incoherent contribution.  All necessary  information on fluxes, efficiencies, POT etc for LSND has been taken from \cite{Aguilar:2001ty} and references therein.

Fig.~\ref{MB-LSND-events} (bottom-left panel) shows our results in comparison to the LSND data for $R_\gamma >10$, where $R_\gamma$ is a parameter defined by the LSND Collaboration (see, for instance \cite{Aguilar:2001ty}) that represents a likelihood ratio that the observed photon signalling the presence of the neutron was correlated as opposed to being accidental. This plot shows the energy distribution and the excess events in the data, as well as those resulting from  our model using the same benchmark parameters as were used to generate the MB results. We find a total of 28.7 events from our model, compared to the 32 events seen by LSND for this choice of $R_\gamma$.

Fig.~\ref{MB-LSND-events} (bottom-right panel)  shows the angular distribution of the light due to the electron-like final state, for $R_\gamma >1$ and visible energies in the range\footnote{The bottom panels use different ranges of $R_\gamma$ and ${\rm E}_{\rm vis}$, because we have chosen to present our results to correspond to the generally available results presented by the LSND Collaboration, which use different $R_\gamma$ and ${\rm E}_{\rm vis}$ ranges for the energy and angular distributions.} $36$~MeV $< {\rm E}_{\rm vis}<60$~MeV. In both panels, the blue shaded region is the result of our model, shown along with backgrounds and data. 

Several points are  pertinent to understanding the results obtained. We discuss them below:
\begin{itemize}
\vspace{-0.2cm}
 \item All LSND events in our scenario stem from the high energy part of their DIF flux, which is kinematically  capable of producing the $N_2$ ($m_{N_2}\simeq 130$ MeV). This flux originates in $\pi^+$'s  created  in proton collisions in the LSND target (the experiment used two different targets over the running period, $\ie$, water and a high-$Z$ material). This leads to a  beam of $\nu_\mu$'s, which interacts in the detector via $\nu_\mu \, {\rm CH}_2\rightarrow n \, N_2 \, X\rightarrow n\, N_1 \, h' \, X\rightarrow N_1 \, \gamma \, e^+e^-  \, X$  (see Fig.~\ref{FD-SP-LSND-MB}). In the final step the photon is the correlated $\gamma$ with an energy of $2.2$ MeV signifying the capture of the neutron by a nucleus. The decays of both $h' $ and $N_2$ are prompt, while $N_1$ is either long-lived and escapes the detector or decays to lighter invisible states.
\vspace{-0.2cm}
\item
In our  scenario, both $H$ and $h'$ act as  mediators  and  contribute to the total cross section. The contribution of $h'$ is much smaller $(\sim 10\%)$ than that of $H$,  since  $\sin \delta \simeq 0.1$. However,  this plays an important role in  producing  the correct angular distribution in MB. In particular, $h'$ is responsible for a coherent contribution which helps sufficiently populate the first ($\ie$, most forward) bin in the top-right panel of Fig.~\ref{MB-LSND-events}.
\vspace{-0.2cm}
\item
 As a consequence of the heavy particle production ($N_2$) necessary, our 
model would not give any signal in KARMEN, which has a narrow-band DIF  flux that peaks at $\sim30$ MeV, hence making it compatible with their null result.
\item The DIF flux, in the oscillation hypothesis, generates electron-like events in energy bins beyond 60 MeV. Indeed, LSND saw $10.5\pm 4.9$ such events (without a correlated neutron) in the range $60$~MeV $<{\rm E}_{\rm vis}<200$~MeV, attributable to an oscillation probability of $(2.9 \pm 1.4)\times10^{-3}$ \cite{Athanassopoulos:1997er}. Our model predicts $34$ such events, which is within their acceptable range of uncertainty.
\item LSND saw about  6 events with a correlated neutron in the energy range $60$~MeV $< {\rm E}_{\rm vis}<200$~MeV, and our calculations yield 5.6 such events, in agreement with their observations.
\vspace{-0.2cm}
\item As mentioned earlier, only  incoherent neutron scattering   contributes to the event counts in LSND. We have assumed  $8$ MeV as the minimum energy transferred to a neutron in order to knock it out and register an event. Additionally,  the masses of $N_2$ and $N_1$ are  important factors in obtaining both the correct number and the correct distributions in this detector. Lowering the mass of $N_2$ increases the total events significantly, since it provides access to lower energies in the DIF flux spectrum.  Decreasing  the  mass of $N_1$ shifts the event peak  towards  higher visible energies, and leads to higher numbers of correlated neutron events with energies $> 60$ MeV, which would conflict with what LSND saw. On the other hand, in MB the effects of $N_2$ and $N_1$ masses do not play as significant a role as they do in LSND, although the MB energy distribution improves if the $N_1$ mass is decreased from our current benchmark value.
\vspace{-0.2cm}
\item Finally, we note that the criteria as to when an $e^+e^-$ pair constitutes a signal that may be counted as an electron-like event in both detectors are different.  MB is not able to distinguish an $e^+ e^-$ pair from a single electron~\cite{1810.07185, Karagiorgi:2010zz} if
the invariant mass of $e^+e^-< 30$~MeV, or if the angle between the pair is $5^\circ$ or less. In our scenario, the mass of $h'$, and hence that of the pair, is $17$~MeV.

In  LSND, the visible energies  are  quite low compared to those in  MB. Hence, the opening angle of the $e^+e^-$ pair can be large for the lower end of the visible energy $(\sim20-30~\rm{MeV})$.  However,  LSND  did not attempt to search for $e^+e^-$ pairs or $\gamma\gamma$ pairs, and for this reason it is reasonable to assume that it  would reconstruct most $e^+e^-$ pairs as a single electron event. In particular,  because timing was their  most powerful particle identifying  variable, the fit to a Cherenkov ring would select the most significant ring, even for large angles between the $e^+e^-$ pair. Therefore, $e^+e^-$ pairs with correlated neutrons would  explain the LSND excess~\cite{William-Louis},
especially since  no known $e^+e^-$ or $\gamma\gamma$ backgrounds were expected in LSND. A more accurate calculation than the simple  one performed here would incorporate the effects of fitting only the most energetic ring out of two which have a large angle between them. One effect of this would be to slightly increase the events in the  middle bin $(36-44$~MeV) at the expense of those at higher energies (including those with energy $> 60$ MeV). It is evident from Fig.~\ref{MB-LSND-events} (bottom-left panel) that this would improve the fit shown.
\end{itemize}
\vspace{-0.7cm}
\subsection{Muon anomalous magnetic moment}
\vspace{-0.3cm}
\label{sec5B}
The one-loop contribution of the light scalars $h',\,H$ to the muon $g-2$ is given by~\cite{Jackiw:1972jz,Leveille:1977rc}
\begin{equation}
\!\!\!\Delta a_{\mu}=\sum_{\phi=h',H}\frac{(y^\phi_{\mu})^2}{8\pi^2}\int_0^1 dx \frac{(1-x)^2(1+x)}{(1-x)^2+x\, (m_\phi/m_\mu)^2},
\end{equation}
where $y^\phi_\mu$ is the coupling strength of $\phi$-$\mu^+$-$\mu^-$ interaction, which is defined in Eq.~(\ref{phi-ll-coupling}).

First, we note that $m_{h'}$, $m_H$ are fixed to fit the LSND and MB measurements. Also, both $h'$ and $H$  contribute to the muon $g-2$, $\Delta a_\mu$ and their ratio $\Delta a_\mu^{h'}/\Delta a_\mu^{H}$ is proportional to $\tan^2\delta$.   In general, $y^\mu$ and the angle $\delta$ would correspond to free parameters and they can be fixed to fit the $\Delta a_\mu$ central value.

For suitably chosen $y^\mu$ and $\delta$ ($y^\mu=1.6\times 10^{-3}$ and $\sin\delta=0.1$), our benchmark yields values which lie in the experimentally allowed $2\sigma$ region, with $\Delta a_\mu\!=\!2.24\times\!10^{-9}$. For these values,  the $H$ contribution to $\Delta a_\mu^H$ is dominant,  with the $h'$ contribution being $16.6\%$ of this quantity.
\vspace{-0.3cm}
\section{Constraints on the model}
\label{sec6}
\vspace{-0.2cm}
This section is devoted to  a discussion of constraints  that the proposed scenario must satisfy, given the couplings of the extended scalar sector to fermions.  We note here that in general, the off-diagonal couplings of the additional scalars  in our model to down/up-type quarks are free parameters and can be very tiny, which is a relevant point that helps us stay safe from several existing bounds, as brought out below.\\
 A second relevant point  in the discussion below  is that we assume that the predominant decay mode for the lightest state among the dark neutrinos $N_i$, $\ie$ $N_1$, is to lighter dark sector particles.

\textit{\bf Constraints from CHARM II and MINER$\nu$A: }
As discussed in~\cite{Arguelles:2018mtc}, these experiments \cite{VILAIN1994246,Valencia:2019mkf} constrain models attempting to explain the MB LEE and LSND based on results of high-energy $\nu$-$e$ scattering. A dark photon ($Z'$) model such as that discussed in~\cite{1807.09877} is tightly constrained for its chosen benchmark values, as shown in~\cite{Arguelles:2018mtc}. We also see that it is possible to evade this constraint provided the value of $|U_{\mu4}|$ (the mixing between the muon and the up-scattered sterile neutrino in the proposed model of \cite{1807.09877})  stays equal to or  below $10^{-4}$. In order to check that our model is safe from these constraints, we calculate the cross section contribution from our process,  (with $H,h'$ as mediators) for CHARM II and MINER$\nu$A and compare its value with that for the model in \cite{1807.09877}, with  $|U_{\mu4}|$ reduced to the safe value of $10^{-4}$. We find that the coherent cross section for our interaction stays more than an order of magnitude below this safe value, comfortably evading this constraint. We note that this is generically true for other recent models with scalar mediators, as also pointed out in \cite{Datta:2020auq,Dutta:2020scq}.

We also note that elastic NC scattering of electrons with $H,h'$ as mediators is not a concern, since the final state contains a $N_2$ which promptly decays to an $h' N_1$ and subsequently a prompt $e^+e^-$. This does not observationally resemble SM $\nu$-$e$ scattering.

\textit{\bf{Constraints from T2K ND280:}} As discussed in~\cite{Brdar:2020tle}, the T2K near-detector, ND280~\cite{T2K:2020lrr}, is in a position to provide bounds on new physics related to the MB LEE. Relevant to our work here, the specific decay $h'\rightarrow e^+e^-$ could be observable in  this detector. Pair production can occur in the Fine-Grained Detectors (FGD), in particular. We have calculated the number of events for our process and find 9 events in FGD1, using a momentum cutoff of 300 MeV and an overall efficiency of 30\%. This is comfortably below their bounds.\\ In principle, at T2K, such events may occur in the TPC also.  In our model, however, this decay is prompt, hence for detection in the TPC, the argon gas in it  must act as both target and detection medium. Since the target mass is only 16 kg, however, the number of events is unobservably small in our case. We also note that the threshold for detection in the TPC is around $200$ MeV.

\textit{\bf{Contributions to NC $\nu$-nucleon scattering at high energies:}}
 Since the $H$ and $h'$ in our model couple to neutrinos and quarks, a possible constraint arises from NC Deep Inelastic Scattering (DIS) of  neutrinos on nucleons, to which these scalars  would contribute as mediators.  At high energies, IceCube and DeepCore are a possible laboratory for new particles which are produced via such  scattering \cite{Coloma:2017ppo, Coloma:2019qqj}. In the process shown in Fig.~\ref{FD-SP-LSND-MB}, the decay  time of $N_2$  (leading to an $e^+e^-$ pair) is short  enough, to escape detection at these detectors. In terms of  distances travelled, this corresponds to  $\sim 1$~m in DeepCore, and $\sim$ a few hundred meters in IceCube, even at very high energies. These lengths are  are much smaller than the  detector resolution necessary to signal a double bang event for both these experiments. In addition, we have checked that  the high energy NC cross section  stays several orders of magnitude below the SM cross section. We also note that $N_1$ in our model  is assumed to decay predominantly to invisible particles, again escaping detection in large detectors.
 
\textit{\bf{ Kaon and $B$-meson decay constraints:}} Prior to discussing specific cases, as a general remark, we note that in any heavy meson decay that involves $u,d$ quarks, one can radiate an $h'$ which would promptly decay to an $e^+e^-$ pair via the diagonal couplings between it and the quarks. While off-diagonal flavour changing couplings in our model are arbitrarily small, the first generation diagonal quark couplings to the scalars in our model are fixed by the requirements of fitting the LSND and MB data, and are approximately ${\cal O}(10^{-5})$. These are small enough to suppress such decays by a factor ${\cal O}(10^{-10})$, rendering them safe from existing upper bounds.
\begin{enumerate}
\item  The BR$(K_L \to \pi^0 e^+ e^-)< 2.8 \times 10^{-10}$ at a 90\% C.L. has been measured at KTeV~\cite{AlaviHarati:2003mr}.
Hence in principle, the width for $K_L \to \pi^0 h'$ would contribute to this, while $K_L\to \pi^0 H$ will not contribute, being kinematically forbidden. However, KTeV applies an invariant mass cut of 140 MeV for the $e^+e^-$ pair, making the bound inapplicable due to kinematics. We note also, as mentioned already,  off-diagonal couplings of $h'$ to $d,s$ quarks in our scenario are tiny. Also, the BR($h'\to\gamma\gamma$) is negligible. Therefore, the constraints from  $K$ decays, $\eg$,  $K_{L,S}\to \pi^0\gamma\gamma$~\cite{Abouzaid:2008xm}  is not applicable.
\vspace{-0.1cm}
\item The E949 Collaboration \cite{Artamonov:2009sz} and NA62 Collaboration~\cite{Ruggiero:2020phq} have measured the process $K^+ \to \pi^+\bar\nu\nu$, which could be mimicked by the $K^+ \to \pi^+ h'$ decay in our scenario. Since $h'$ decays
primarily to $e^+e^-$, this means that it must be long-lived and escape the
detector for this bound to apply. From~\cite{Liu:2020qgx}, we see
that given $h'$ mass of 17~MeV in our model,  as long as lifetime is less
than  approximately $10^{-10}$~s, one is safe from the constraint from
invisibles. (In our model, $h'$ has a lifetime $\simeq 1.8\times 10^{-12}$~s.) Moreover, as mentioned above,  off-diagonal couplings of $h'$ to $d,s$ quarks in our scenario are tiny. 
\vspace{-0.1cm}
\item A  light scalar coupled to muons can be emitted in the decay $K^+\to\mu^+ \nu\phi$.  This is constrained by the ~NA62~\cite{Martellotti:2015kna}, as discussed in~\cite{Batell:2016ove}.  $H$ will evade these constraints because of its large mass, while $h'$ lies outside the constrained range due to its short lifetime and small coupling to muons. In addition, we note that data collected by the NA48/2 experiment~\cite{Madigozhin:2019tcs} can in principle provide constraints via observation of $K^+\to\mu^+ \nu e^+e^-$, as noted in~\cite{Batell:2016ove}. This analysis has recently been done~\cite{Madigozhin:2019tcs}, but with an invariant minimum mass cut of $140$ MeV for the $e^+e^-$ pair, which makes it inapplicable to $h'$.
\vspace{-0.1cm}
 \item The CHARM experiment~\cite{CHARM:1983ayi,Bergsma:1985is} has measured the displaced decay of neutral particles into $\gamma\gamma,\, e^+e^-$ and $\mu^+\mu^-$. The relevant decays are  $K_L \to \pi^0 h'$ and $K^+ \to \pi^+ h'$.  Thus in our model,  $h'$ can in principle be  constrained by this experiment, but  as discussed in~\cite{Liu:2020qgx}, for $m_{h'}\simeq 17$~MeV,  the lifetime in our case is much shorter than $10^{-10}$,
which is the upper value set by this bound. Additionally, it is possible that CHARM, being sensitive to heavy neutral leptons given its dimensions,  could have  sensitivity to visible decays of  $N_1$. As noted earlier, however, $N_1$ decays primarily to invisible states.
\vspace{-0.1cm}
\item The $K_{\mu2}$ experiment~\cite{Yamazaki:1984vg} has measured  the $K^+ \to \pi^+\phi$ process. For our benchmark point (Table~\ref{tab}), BR$(K^+ \to \pi^+h')\simeq 4.2\times 10^{-12}$~\cite{Batell:2018fqo}, which is very small compared with the upper limit $\sim 10^{-8}$. 
\vspace{-0.1cm}
\item Similarly, the decay $B\to K^*e^+e^-$ has been measured at the LHCb~\cite{Aaij:2013hha}, ${\rm BR}(B \to K^* e^+ e^-) = (4.2\pm0.7)\times10^{-7}$. In our model, this would correspond to $B\to K^* h'/H$, with the latter going to $e^+e^-$. Given  the $b\to s$ transition involved,  and  that  couplings of $H, h'$ to $b, s$ quarks can be arbitrarily small, we evade this constraint. Also, due to $2 m_e<m_H<2m_\tau$, the decays $B\to K^{(*)} H\to  K^{(*)} \mu^+ \mu^-$ are subjected to strong constraints from $B\to K^{(*)} \mu^+ \mu^-$ at the LHCb~\cite{Aaij:2015tna}. However, in our model, we evade these constraints because of the smallness of $H$ coupling to $b, s$ quarks. Finally, in our model, it is worth mentioning that the branching ratios of $B\to K^{(*)}\gamma\gamma/\nu\bar\nu$ are negligible.
\vspace{-0.15cm}
\end{enumerate}
{\textit{\bf Constraints from neutrino trident production:}} The neutrino trident process \cite{Ge:2017poy} provides a sensitive probe of BSM  physics at neutrino detectors, and has been measured \cite{Geiregat:1990gz,Mishra:1991bv, Adams:1998yf}. It is relevant to our model given the couplings of the $H,h'$ to muons, which are used for our explanation to the muon $g-2$ anomaly. Using the SM cross section and simple scaling, we have checked that our model is safe from this constraint.

{\textit{\bf Pion decay constraints:}} $H,h'$  couple to quarks, hence $H,h'$  can mediate $\pi^0$ decay to $e^+e^-$. In the SM, this decay is loop-suppressed and consequently small. However, given the small couplings of the two scalars to the $u,d$ quarks ($\sim 10^{-5}$) and the electron ($\sim 10^{-4}$), we find that we are safely below this constraint.

{\textit{\bf Collider bounds:}} At hadron colliders, the process $Z\to4\ell$ proceeds via $q\bar q \to
Z^*\to \bar\ell \ell$, along with a $\gamma^*$ attached to one of the
external legs (either the quarks or leptons) and with the  creation of a lepton pair
from the~$\gamma^*$.  This process has been measured at the LHC \cite{Rainbolt:2018axw}. The $Z$ and $\gamma$ in principle can be replaced by $H,h'$. The bound from LHC, however,
applies an $\ell^+\ell^-$ invariant mass cut of 4~GeV, and hence  does
not apply to our situation.

Since $H$ couples to leptons and quarks, $H$ can be radiated from these in
external and internal legs in any process, and then $H\to h' h'$ is
possible, leading to two collimated pairs of $e^+e^-$, which will look like two
leptons. An LHC search was conducted \cite{Aad:2015sms} and no significant
deviation or excess was found. As discussed in \cite{Liu:2020qgx}, given the fact that 
the $H$ to lepton/quark coupling is small ($\lesssim 10^{-4}$), and also that the $H\to h'h'$ decay width is small due to the smallness of its coupling, the contribution  will stay within the 1\% level.

If $H, h'$ couple to $b$ and $s$ quarks, then the decay $B_s \to \mu^+\mu^-$ can
be mediated  by them. This decay has been measured by both the LHCb and CMS (see
\cite{Aaij:2013aka, Chatrchyan:2013bka, CMS:2014xfa, Aaij:2017vad, Aaboud:2018mst}). However,  in our model the couplings to $b$, $s$ quarks can be arbitrarily small, hence  this constraint can be avoided. 

\textit{\bf{Constraints on $y_e$ and $m_{h'}$ from dark photon searches:}} A generic dark photon search looks for its decay to a  pair of leptons. One may translate such bounds ~\cite{Alves:2017avw, Knapen:2017xzo} to constraints on a light scalar with couplings  to leptons. Specifically,  translated constraints relevant to our scenario arise from KLOE \cite{Anastasi:2015qla} and BABAR~\cite{Lees:2014xha}. Current values of $y_e$ and $m_{h'}$ in our scenario are safe from these bounds, but they will be tested in the future by  Belle-II~\cite{Batell:2017kty}.

\textit{\bf{Constraints on $y_e$ and $m_{h'}$ from electron beam dump experiments:}} As discussed in ~\cite{Liu:2016qwd, Batell:2016ove}, a light scalar with couplings  to electrons could be detected in beam dump experiments if it decays  to an $e^+ e^-$ pair or to photons. For  the mass range relevant here,   the experiments E137~\cite{Bjorken:1988as}, E141~\cite{Riordan:1987aw}, ORSAY~\cite{Davier:1989wz}   and NA64~\cite{Banerjee:2018vgk} can potentially provide restrictive bounds.  While our present values are outside the  forbidden regions,  they will be tested  in the future by  the HPS fixed target experiment \cite{Battaglieri:2014hga} which will scatter electrons on tungsten. 

\textit{\bf{Constraints on $y_e$ and $m_{H}$ from dark photon searches:}} KLOE~\cite{Anastasi:2018azp} searched for $e^+e^- \to U\gamma $, followed by  $U$ decays to $\pi^+\pi^-/\mu^+\mu^-$ leading to the constraint on $y_e\, (\sim 2\times 10^{-4})$ at $m_{U}=750$~MeV.  In our scenario, replacing $U$ by $H$ we note that the production of $\pi^+\pi^-/\mu^+\mu^-$ by it in KLOE will be very suppressed due to its tiny coupling to $u,d$ quarks and its predominant semi-visible decay $(H\to e^+e^-+\slashed{E})$. Moreover, both visible and invisible final states searches by BABAR~\cite{Lees:2014xha,Lees:2017lec} put upper limits on $y_e$ at $m_H$, which can be evaded by $H$ due to its predominant semi-visible decay.

\textit{\bf{Constraints on $y_\mu$ and $m_H$ from colliders:}} BABAR has provided constraints \cite{Batell:2016ove, Batell:2017kty} on these parameters via their search for $e^+e^-\rightarrow \mu^+\mu^-\phi$, where $\phi$ is a generic light scalar. Our values, while currently in conformity~with these bounds, will be tested in the future by Belle-II~\cite{Batell:2017kty}. In addition, BABAR~\cite{BABAR:2020oaf} constrains a dark leptophillic light scalar with couplings which are proportional to~$m_f/v$. This set of constraints does not apply in our case since our couplings for $h'$ and $H$ do not have this proportionality.

\textit{\bf{Contribution from the new scalars to the electron $g-2$ anomaly:}} The (positive) one  loop contribution in our model allows us to explain the observed value of $\Delta a_{\mu}$. A similar (positive) contribution is made to $\Delta a_{e}$ by both $h'$ and $H$, which we have computed, summed and found to be $\Delta a_{e} = 4 \times 10^{-14}$. This is well within the present uncertainties in this quantity. We note that our model allows the possibility of having negative off-diagonal Yukawa couplings to  $\Delta a_{e}$ by both $h'$ and $H$. This affords flexibility in varying this contribution and keeping it within acceptable limits, as well as possibly explaining the current  $\Delta a_{e}$ discrepancy at the one-loop level, as discussed in~\cite{Dutta:2020scq}.
\vspace{-0.7cm}
\section{Conclusions}
\label{sec7}
\vspace{-0.4cm}
Evidence for anomalous signals at low-energy non-collider experiments in general,  and short baseline neutrino experiments in particular, has been gradually increasing over time, and has firmed up significantly over  the past decade or so. Specifically, with reference to the LSND and MB excesses, it has gradually become evident that one may choose several different approaches  towards understanding their origin, and these choices can  lead  down divergent and non-overlapping paths\footnote{For example, 
one could conclude that these anomalies are due to poorly understood  SM backgrounds or detector-specific systematic effects, as opposed to trying to understand them via active-sterile neutrino oscillations. Clearly, these choices would subsequently  entail very different theoretical and experimental efforts towards their eventual resolution.}.

 An important premise underlying our effort in this paper is that a common, non-oscillation, new physics explanation exists for both LSND and MB. Furthermore, our effort is guided by the belief that such an explanation could not only yield a long-sought extension to the SM, but also delineate the contours of the portal connecting the SM to the dark sector, as well as shed some light on other related but as yet unresolved  questions.

 Pursuant to this, in the scenario presented here, the extension to the SM that these experiments lead to comprises of the well-known 2HDM. Access to the dark sector is achieved via mass-mixing with a  (dark) relatively light singlet scalar and via the presence of heavier dark neutrinos in allowed gauge invariant terms in the Lagrangian. Two of the three CP-even scalars in the model are relatively light ($m_{h'}\simeq 17$~MeV and $m_H\simeq 750$~MeV) and participate in the interaction that generates the excesses in LSND and MB, as well as contribute to the value of the muon $g-2$. Similarly, two of the three  dark neutrinos not only participate in important ways in the interaction\footnote{The third dark neutrino, $N_3$, with $m_{N_3} \simeq 10$ GeV does not participate because of kinematics, not dynamics.} in LSND and MB,  but also, along with the third neutrino, generate neutrino masses via a simple seesaw mechanism. 
 
The sub-GeV scalars in our model can be searched for in a variety of experiments. The masses of $h'$ and $H$ lie especially close to existing bounds from electron beam-dump experiments like E141~\cite{Riordan:1987aw} and BABAR~\cite{Lees:2014xha} respectively. Thus $H$ can be searched for in  Belle-II~\cite{Batell:2017kty}  and $h'$ in  HPS~\cite{Battaglieri:2014hga}. The dark fermions in our model are amenable to searches in several upcoming experiments, $\eg$  DUNE ND~\cite{Adams:2013qkq} (for a more detailed discussion and references, see \cite{Ballett:2019bgd,Berryman:2019dme}). 

In the near future, the MicroBooNE experiment \cite{Caratelli:2020ues, Foppiani:2019sgs,Kaleko:2017bgx} will provide first indications of whether the low-energy electron-like event excesses in MB and LSND are due to electrons or photons. In the scenario presented here, the $h'$ has a very short lifetime prior to decay to an $e^+e^-$ pair. At the energies under consideration, it would travel about $5-12$~mm in the detector. Since tracks with a gap of greater than $1$~cm would be interpreted as photons, most events resulting from our scenario would look like an excess of electrons in MicroBooNE, while the high energy ones could be mistaken for photons with short gaps. A $dE/dx$ analysis would be required to actually detect that the events are $e^+e^-$ pairs rather than electrons, which should also be possible with more data.

 With respect to the scalar search for the $ h'$, we mention two existing  experimental hints which are interesting: $a)$ a significant excess in the $10-20$ MeV invariant mass-bin of electron-like FGD1-TPC pairs detected by the T2K ND280 detector, (see Fig.~11 in~\cite{T2K:2020lrr}), and $b)$ the higher than expected central value for the width $\Gamma(\pi^0\rightarrow e^+e^-)$ observed by the KTeV experiment~\cite{Abouzaid:2006kk}, signifying the possible existence of a scalar with mass $\simeq 17$ MeV. Additionally, we would like to point out that the Kaon DAR search,  planned at the JSNS$^2$ experiment \cite{Ajimura:2017fld, Jordan:Poster-Neutrino-2020} is in a position to provide a test of the proposal presented in this work via its flux of high energy $\nu_\mu$ \cite{William-Louis-thanks}.

In conclusion, we are hopeful that the long-standing and statistically significant  anomalous results of LSND and MB, along with the connection established between them via the simple model presented here will help motivate a more focused search for these particles in ongoing and future experiments.

\section*{Acknowledgements} 
RG would like to express his sincere appreciation to William Louis for his help with our many questions on MB and LSND.  He is thankful to Boris Kayser and Geralyn Zeller for helpful discussions, and acknowledges useful  email exchanges with Lukas Koch, Teppei Katori and Zoya Vallari. He is grateful to  the Theory Division and the 
 Neutrino Physics Center at Fermilab for  hospitality and visits where this work benefitted from discussions and a conducive environment. SR is grateful to Fermilab for support via the Rajendran Raja Fellowship.  WA, RG and SR also acknowledge support from the XII Plan Neutrino Project of the Department of Atomic Energy and the High Performance Cluster Facility at HRI (http://www.hri.res.in/cluster/). 
\bibliographystyle{apsrev}
\bibliography{NU-bib}
\end{document}